\documentclass{article}
\usepackage[T1]{fontenc} \usepackage[utf8]{inputenc}
\usepackage[svgnames]{xcolor}
\usepackage[bookmarks=false,colorlinks=true,allcolors=DarkBlue]{hyperref} 
\usepackage{ismir,amsmath}
\usepackage{graphicx} \graphicspath{{figs/}}
\usepackage[font=small,skip=8pt]{caption}
\usepackage{booktabs, threeparttable, tabularx}
\usepackage{enumitem} \setlist{itemsep=0pt,topsep=0pt,parsep=0pt,partopsep=0pt}
\usepackage{scrextend}

\newcommand{\todo}[1]{{\color{red} #1}}

\newcommand{\ntracks}{106,574 }
\newcommand{\nartists}{16,341 }
\newcommand{\nalbums}{14,854 }
\newcommand{\ngenres}{161 }
\newcommand{\tduration}{343 }
\newcommand{\aduration}{278 }
\newcommand{\size}{917 }
\newcommand{\weblink}{https://github.com/mdeff/fma}

\newcommand{\tspacea}{\hspace{0.7em}}
\newcommand{\tspaceb}{\hspace{0.9em}}

\title{FMA: A Dataset For Music Analysis}

\multauthor
{Michaël Defferrard$^\dagger$ \hspace{.6cm} Kirell Benzi$^\dagger$ \hspace{.6cm} Pierre Vandergheynst$^\dagger$ \hspace{.6cm} Xavier Bresson$^\ddagger$}
{
	$^\dagger$LTS2, EPFL, Switzerland \hspace{0.4cm}
	$^\ddagger$SCSE, NTU, Singapore \hspace{0.4cm}
	{\small Work conducted when XB was at EPFL.}\\
	{\tt\small \{michael.defferrard,kirell.benzi,pierre.vandergheynst\}@epfl.ch, xbresson@ntu.edu.sg}
}
\def\authorname{Michaël Defferrard, Kirell Benzi, Pierre Vandergheynst, Xavier Bresson}

\begin{document}
\maketitle

\begin{abstract}
We introduce the Free Music Archive (FMA), an open and easily accessible dataset suitable for evaluating several tasks in MIR, a field concerned with browsing, searching, and organizing large music collections.
The community's growing interest in feature and end-to-end learning is however restrained by the limited availability of large audio datasets.
The FMA aims to overcome this hurdle by providing \size GiB and \tduration days of Creative Commons-licensed audio from \ntracks tracks from \nartists artists and \nalbums albums, arranged in a hierarchical taxonomy of \ngenres genres.
It provides full-length and high-quality audio, pre-computed features, together with track- and user-level metadata, tags, and free-form text such as biographies.
We here describe the dataset and how it was created, propose a train/validation/test split and three subsets, discuss some suitable MIR tasks, and evaluate some
baselines for genre recognition.
Code, data, and usage examples are available at \url{\weblink}.
\end{abstract}

\section{Introduction} 

While the development of new mathematical models and algorithms to solve challenging real-world problems is obviously of first importance to any field of research, evaluation and comparison to the existing state-of-the-art is necessary for a technique to be widely adopted by research communities. Such tasks require open benchmark datasets to be reproducible. 
In computer vision, the community has developed established benchmark datasets such as \href{http://yann.lecun.com/exdb/mnist/}{MNIST}~\cite{mnist}, \href{https://www.cs.toronto.edu/~kriz/cifar.html}{CIFAR}~\cite{cifar}, or \href{http://www.image-net.org}{ImageNet}~\cite{imagenet}, which have proved essential to advance the field. The most celebrated example, the \href{http://www.image-net.org/challenges/LSVRC/2012/}{ILSVRC2012 challenge} on an unprecedented ImageNet subset of 1.3M images~\cite{imagenet_challenge}, demonstrated the power of deep learning (DL), which won the competition with an 11\% accuracy advantage over the second best~\cite{convnet_imagenet}, and enabled incredible achievements in both fields~\cite{dl}.

\begin{table}[t]
	\small
	\centering
	\begin{threeparttable}
	\begin{tabular}{l@{ }rrcc}
		\toprule
		dataset\tnote{1} & \#clips & \#artists & year & audio \\
		\midrule
		\href{https://staff.aist.go.jp/m.goto/RWC-MDB/}{RWC}~\cite{RWC} & 465 & - & 2001 & yes \\
		\href{http://calab1.ucsd.edu/~datasets/cal500/}{CAL500}~\cite{cal500} & 500 & 500 & 2007 & yes \\
		\href{http://mtg.upf.edu/ismir2004/contest/tempoContest/node5.html}{Ballroom}~\cite{ballroom} & 698 & - & 2004 & yes \\
		\href{https://marsyasweb.appspot.com/download/data_sets/}{GTZAN}~\cite{gtzan} & 1,000 & $\sim300$ & 2002 & yes \\
		\href{http://www.cp.jku.at/datasets/musiclef/}{MusiClef}~\cite{musiclef} & 1,355 & 218 & 2012 & yes \\
		\href{https://labrosa.ee.columbia.edu/projects/artistid/}{Artist20}~\cite{artist20} & 1,413 & 20 & 2007 & yes \\
		\href{http://ismir2004.ismir.net/genre_contest/}{ISMIR2004} & 1,458 & - & 2004 & yes \\
		\href{http://www-ai.cs.uni-dortmund.de/audio.html}{Homburg}~\cite{homburg} & 1,886 & 1,463 & 2005 & yes \\  
		103-Artists~\cite{103artists} & 2,445 & 103 & 2005 & yes \\
		\href{http://www.seyerlehner.info/index.php?p=1_3_Download}{Unique}~\cite{unique} & 3,115 & 3,115 & 2010 & yes \\
		\href{http://www.seyerlehner.info/index.php?p=1_3_Download}{1517-Artists}~\cite{1517artists} & 3,180 & 1,517 & 2008 & yes \\
		\href{http://www.ppgia.pucpr.br/~silla/lmd/}{LMD}~\cite{lmd} & 3,227 & - & 2007 & no \\
		\href{http://anasynth.ircam.fr/home/media/ExtendedBallroom}{EBallroom}~\cite{extballroom} & 4,180 & - & 2016 & no\tnote{2} \\
		\href{https://labrosa.ee.columbia.edu/projects/musicsim/uspop2002.html}{USPOP}~\cite{uspop} & 8,752 & 400 & 2003 & no \\
		\href{http://calab1.ucsd.edu/~datasets/cal10k/}{CAL10k}~\cite{cal10k} & 10,271 & 4,597 & 2010 & no \\ 
		\href{http://mirg.city.ac.uk/codeapps/the-magnatagatune-dataset}{MagnaTagATune}~\cite{magnatagatune} & 25,863\tnote{3} & 230 & 2009 & yes\tnote{4} \\ 
		\href{http://jmir.sourceforge.net/index_Codaich.html}{Codaich}~\cite{codaich} & 26,420 & 1,941 & 2006 & no \\ 
		\bf \href{\weblink}{FMA} & \bf \ntracks & \bf \nartists & \bf 2017 & \bf yes \\
		\href{http://www.omras2.org/}{OMRAS2}~\cite{omras} & 152,410 & 6,938 & 2009 & no \\
		\href{https://labrosa.ee.columbia.edu/millionsong/}{MSD}~\cite{msd} & 1,000,000 & 44,745 & 2011 & no\tnote{2} \\
		\href{https://research.google.com/audioset/}{AudioSet}~\cite{audioset} & 2,084,320 & - & 2017 & no\tnote{2} \\
		\href{https://acousticbrainz.org}{AcousticBrainz}~\cite{acousticbrainz} & 2,524,739\tnote{5} & - & 2017 & no \\
		\bottomrule
	\end{tabular}
	\begin{tablenotes}
		\item[1] Names are clickable links to datasets' homepage.
		\item[2] Audio not directly available, can be downloaded from \\ \href{http://www.ballroomdancers.com}{ballroomdancers.com}, \href{https://www.7digital.com}{7digital.com}, \href{https://www.youtube.com}{youtube.com}.
		\item[3] The 25,863 clips are cut from 5,405 songs.
		\item[4] Low quality 16 kHz, 32 kbit/s, mono mp3.
		\item[5] As of 2017-07-14, of which a subset has been linked to genre labels for the \href{http://www.multimediaeval.org/mediaeval2017}{MediaEval 2017} \href{https://multimediaeval.github.io/2017-AcousticBrainz-Genre-Task}{genre task}.
	\end{tablenotes}
	\end{threeparttable}
	\caption{Comparison between FMA and alternative datasets.}
	\label{tab:datasets}
\end{table}

\begin{table}
	\small
	\centering
	\begin{threeparttable}
	\begin{tabular}{|r@{ }l@{\hspace{.8em}}r@{ }l@{\hspace{.8em}}r@{ }l|}
		\hline
		100\% & track\_id     & 100\% & title          &  93\% & number \\
		  2\% & information   &  14\% & language\_code & 100\% & license \\
		  4\% & composer      &   1\% & publisher      &   1\% & lyricist \\
		 98\% & genres        &  98\% & genres\_all    &  47\% & genre\_top \\
		100\% & duration      & 100\% & bit\_rate      & 100\% & interest \\ 
		100\% & \#listens     &   2\% & \#comments     &  61\% & \#favorites \\
		100\% & date\_created &   6\% & date\_recorded &  22\% & tags\\
		\hline
		100\% & album\_id     & 100\% & title          &     &  \\
		 94\% & type          &  96\% & \#tracks       &     &  \\
		 76\% & information   &  16\% & engineer       &  18\% & producer \\
		 97\% & \#listens     &  12\% & \#comments     &  38\% & \#favorites \\
		 97\% & date\_created &  64\% & date\_released &  18\% & tags \\
		\hline
		100\% & artist\_id    & 100\% & name           &  25\% & members \\
		 38\% & bio           &   5\% & \multicolumn{2}{@{}l}{associated\_labels} &  \\
		 43\% & website       &   2\% & wikipedia\_page & &  \\
		& & 5\% & \multicolumn{2}{@{}l}{related\_projects} & \\
		 37\% & location      &  23\% & longitude      &  23\% & latitude \\
		 11\% & \#comments    &  48\% & \#favorites    &  10\% & tags\tnote{1} \\
		 99\% & date\_created &   8\% & active\_year\_begin & & \\
		& & 2\% & \multicolumn{2}{@{}l}{active\_year\_end} & \\
		\hline
	\end{tabular}
	\begin{tablenotes}
		\item[1] One of the tags is often the artist name. It has been subtracted.
	\end{tablenotes}
	\end{threeparttable}
	\caption{List of available per-track, per-album and per-artist metadata, i.e.\ the columns of \texttt{tracks.csv}. Percentages indicate coverage over all tracks, albums, and artists.}
	\label{tab:metadata}
\end{table}

Unlike the wealth of available visual or textual content, the lack of a large, complete and easily available dataset for MIR has hindered research on data-heavy models such as DL.
\tabref{tab:datasets} lists the most common datasets used for content-based MIR.
GTZAN~\cite{gtzan}, a collection of 1000 clips from 10 genres, was the first publicly available benchmark dataset for genre recognition (MGR). As a result, despites its flaws (mislabeling, repetitions, and distortions), it continues to be the most used dataset for MGR~\cite{mgr_eval_2}. Moreover, it is small and misses metadata which e.g.\ prevents researchers to control for artists or album effects. 
Looking at \tabref{tab:datasets}, the well-known MagnaTagATune~\cite{magnatagatune} and the Million Song Dataset (MSD)~\cite{msd} as well as the newer AudioSet~\cite{audioset} and AcousticBrainz~\cite{acousticbrainz} appear as contenders for a large-scale reference dataset.
MagnaTagATune, which was collected from the \href{https://magnatune.com/}{Magnatune} label and tagged using the \href{http://tagatune.org/}{TagATune} game, includes metadata, features and audio. The poor audio quality and limited number of songs does however limit its usage.
MSD and AudioSet, while very large, force researchers to download audio clips from online services. AcousticBrainz's approach to the copyright issue is to ask the community to upload music descriptors of their tracks. Although it is the largest database to date, it will never distribute audio.
On the other hand, the proposed dataset offers the following qualities, which in our view are essential for a reference benchmark.



\textbf{Large scale.} Large datasets are needed to avoid overtraining and to effectively learn models that incorporate the ambiguities and inconsistencies that one finds with musical categories. They are also more diverse and allows to average out annotation noise as well as characteristics who might be confounded with the ground truth and exploited by learning algorithms.
While FMA features less clips than MSD or AudioSet, every other dataset with available quality audio are two orders of magnitude smaller (\tabref{tab:datasets}).

\textbf{Permissive licensing.} MIR research has historically suffered from the lack of publicly available benchmark datasets, which stem from the commercial interest in music by record labels, and therefore imposed rigid copyright.
The FMA's solution is to aim for tracks which license permits redistribution.
All data and code produced by ourselves are licensed under the \href{https://creativecommons.org/licenses/by/4.0)}{CC BY 4.0} and MIT licenses.

\textbf{Available audio.}
\tabref{tab:datasets} shows that while the smaller datasets are usually distributed with audio, most of the larger do not. They either (i) only contain features derived from the audio, or (ii) provide links to download the audio from an online service.\footnote{Going to the source distributor is a way to adhere with copyright.}
The problem with (i) is that researchers are stuck with the chosen features and are prevented to leverage feature learning or end-to-end learning systems like DL. Moreover, we should be wary of proprietary features like those computed by commercial services such as \href{http://the.echonest.com/}{Echonest}.
The problem with (ii) is that researchers have no control, i.e.\ we have no assurance that the files or services will not disappear or change without notice.

\textbf{Quality audio.} 
Distributed or downloadable audio are usually clips of 10 to 30 seconds and sometimes of low quality, e.g.\ 32 kbit/s for MagnaTagATune or an average of 104 kbit/s for MSD~\cite{msd_features}. The problem with clips is that the beginning 30 seconds of tracks may yield different results than the middle or final 30 seconds, and that researchers may not have control over which part they get.
In comparison, FMA comes with full-length and high-quality audio.

\textbf{Metadata rich.} The dataset comes with rich metadata, shown in \tabref{tab:metadata}. While not complete in any means, it compares favorably with the MSD which only provides artist-level metadata~\cite{msd} or GTZAN which offers none.

\textbf{Easily accessible.} Working with the dataset only requires to download some \texttt{.zip} archives containing \texttt{.csv} metadata and \texttt{.mp3} audio. No need to crawl the web and circumvent rate limits or access restrictions. Besides, we provide
some usage examples in the \texttt{usage.ipynb} Jupyter notebook to start using the data quickly.

\textbf{Future proof and reproducible.} All files and archives are checksummed and hosted in a long-term digital archive.
Doing so alleviates the risks of songs to become unavailable.
Moreover, we share all the code used to (i) collect the data, (ii) analyze it, (iii) generate the subsets and splits, (iv) compute the features and (v) test the baselines. The developed code can serve as a starting point for researchers to compute their own features or evaluate their methods.
Finally, anybody can recreate or extend the collection, thanks to public songs and APIs.

Note that an alternative to open benchmarking is the approach taken by the MIREX evaluation challenges: the evaluation (by the organizers) of submitted algorithms on private datasets~\cite{mirex}. This practice however incurs an approximately linear cost with the number of submissions, which put the long-term sustainability of MIREX at risk~\cite{mirex_critic}. By releasing this open dataset, we realize part of the vision of McFee \textit{et al}.\ in ``a distributed, community-centric paradigm for system evaluation, built upon the principles of openness, transparency, and reproducibility''.

\begin{figure*}
	\begin{minipage}{0.74\linewidth}
		\centering
		\includegraphics[width=\linewidth]{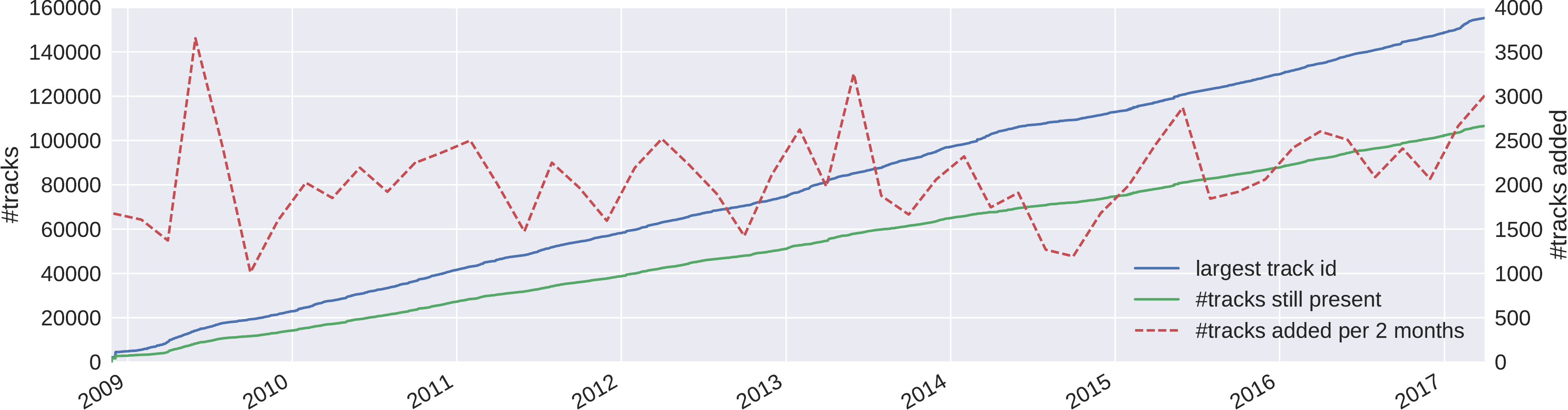}
	\end{minipage} \hfill
	\begin{minipage}{0.24\linewidth}
		\centering
		\includegraphics[width=\linewidth]{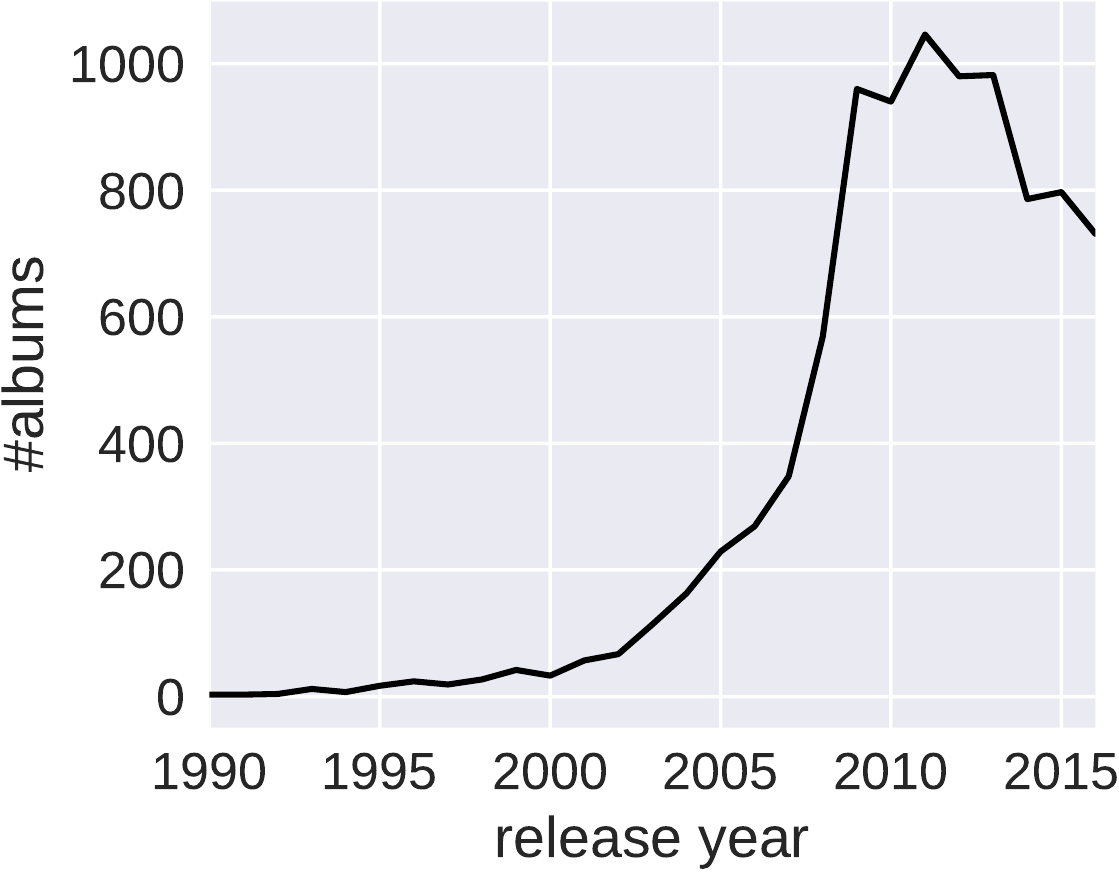}
	\end{minipage}
	\caption{(left) Growth of the archive, created in 11/2008. (right) Number of albums released per year (min 1902, max 2017).}
	\label{fig:growth}
	\label{fig:album_release_year}
\end{figure*}

\begin{table*}
	\tiny
	\centering
\begin{tabular}{@{}l@{\tspacea}l@{\tspacea}l@{\tspacea}l@{\tspacea}r@{\tspacea}r@{\tspacea}l@{\tspacea}r@{\tspacea}l@{\tspacea}l@{\tspacea}l@{}}
	\toprule
	\multicolumn{6}{@{}l}{track} & \multicolumn{3}{@{}l}{album} & \multicolumn{2}{@{}l}{artist} \\
	track\_id & title & genres\_all & genre\_top & dur. & listens & title & listens & tags & name & location \\
	\midrule
	150073 & Welcome to Asia & [2, 79] & International & 81 & 683 & Reprise & 4091 & [world music, dubtronica, fusion] & DubRaJah & Russia \\
	140943 & Sleepless Nights & [322, 5] & Classical & 246 & 1777 & Creative Commons Vol. 7 & 28900 & [classical, alternate, soundtrack, piano, ... & Dexter Britain & United Kingdom \\
	64604 & i dont want to die alone & [32, 38, 456] & Experimental & 138 & 830 & Summer Gut String & 7408 & [improvised, minimalist, noise, ... & Buildings and Mountains & Oneonta, NY \\
	23500 & A Life In A Day & [236, 286, 15] & Electronic & 264 & 1149 & A Life in a Day & 6691 & [idm, candlestick, romanian, candle, ... & Candlestickmaker & Romania \\
	131150 & Yeti-Bo-Betty & [25, 12, 85] & Rock & 124 & 183 & No Life After Crypts & 3594 & [richmond, fredericksburg, trash rock, ... & The Crypts! & Fredericksburg \\
	\bottomrule
	\end{tabular}
	\caption{Some rows and columns of the metadata table, stored in \texttt{tracks.csv}.}
	\label{tab:tracks}
\end{table*}

\section{Dataset} 

\subsection{The Free Music Archive}

The dataset, both the audio and metadata, is a dump of the \href{https://freemusicarchive.org/}{Free Music Archive}, a free and open library directed by \href{https://wfmu.org/}{WFMU}, the longest-running freeform radio station in the United States.
Inspired by \href{https://creativecommons.org/}{Creative Commons} and the open-source software movement, the FMA provides a platform for curators, artists, and listeners to harness the potential of music sharing.
The website provides a large catalog of artists and tracks, hand-picked by established audio curators. Each track is legally free to download as artists decided to release their works under permissive licenses.
While there exists other sources of CC-licensed music, notably \href{https://www.jamendo.com}{Jamendo}, FMA is unique as it combines user-generated content with the curatorial role that WFMU and others have always played.\footnote{\href{http://rhizome.org/editorial/2009/may/1/interview-with-jason-sigal-of-the-free-music-archi}{Interview with Jason Sigal of the Free Music Archive, Rhizome.}}

\subsection{Creation} 

As of April 1st 2017, when the dataset was gathered, the online archive largest track id was 155,320, of which 109,727 were valid. The missing 45,594 ids probably correspond to deleted tracks. \figref{fig:growth} illustrates the growth of the dataset. In addition to per-track metadata, the used hierarchy of 161 genres and extended per-album (480 not found) and per-artist (250 not found) metadata were collected via the available \href{https://freemusicarchive.org/api}{API}.\footnote{See \texttt{webapi.ipynb} to query the API with our helpers.} Finally, mp3 audio was downloaded over HTTPS.
Out of all collected track ids, 180 mp3s could not be downloaded, 286 could not be trimmed by ffmpeg, and features could not be extracted from 71. Finally, the license of 2,616 tracks prohibited their redistribution, leaving us with \ntracks tracks.

While it may be argued that the dataset should be cleaned, we wanted it to resemble real world data. As such, we did not remove tracks which have too many genres, are too long, belong to rare genres, etc. Moreover, it is hard to set a threshold, algorithms shall handle outliers, and the small number of outliers will not impact performance much anyway. Researchers are obviously free to discard any track for training.


\begin{figure}
	\centering
	\includegraphics[width=\linewidth]{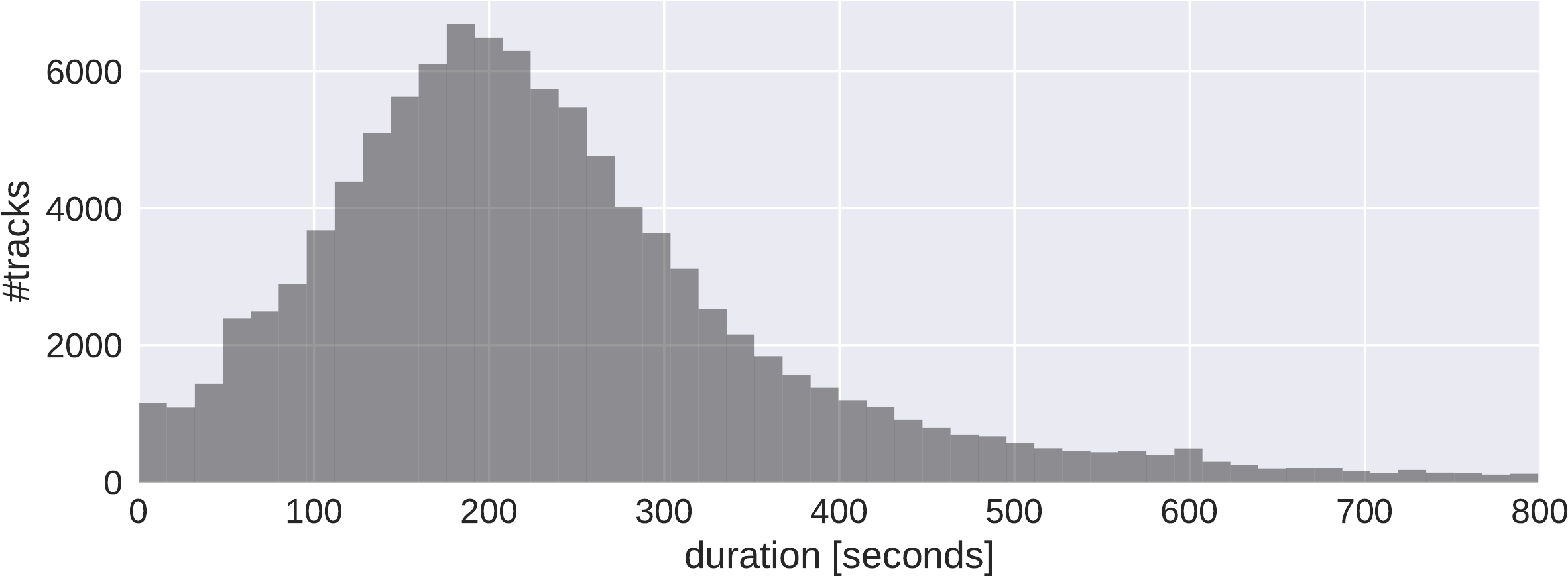}
	\caption{Track duration (min 0, max 3 hours).}
	\label{fig:duration_distribution}
\end{figure}

\begin{figure}
	\centering
	\includegraphics[width=\linewidth]{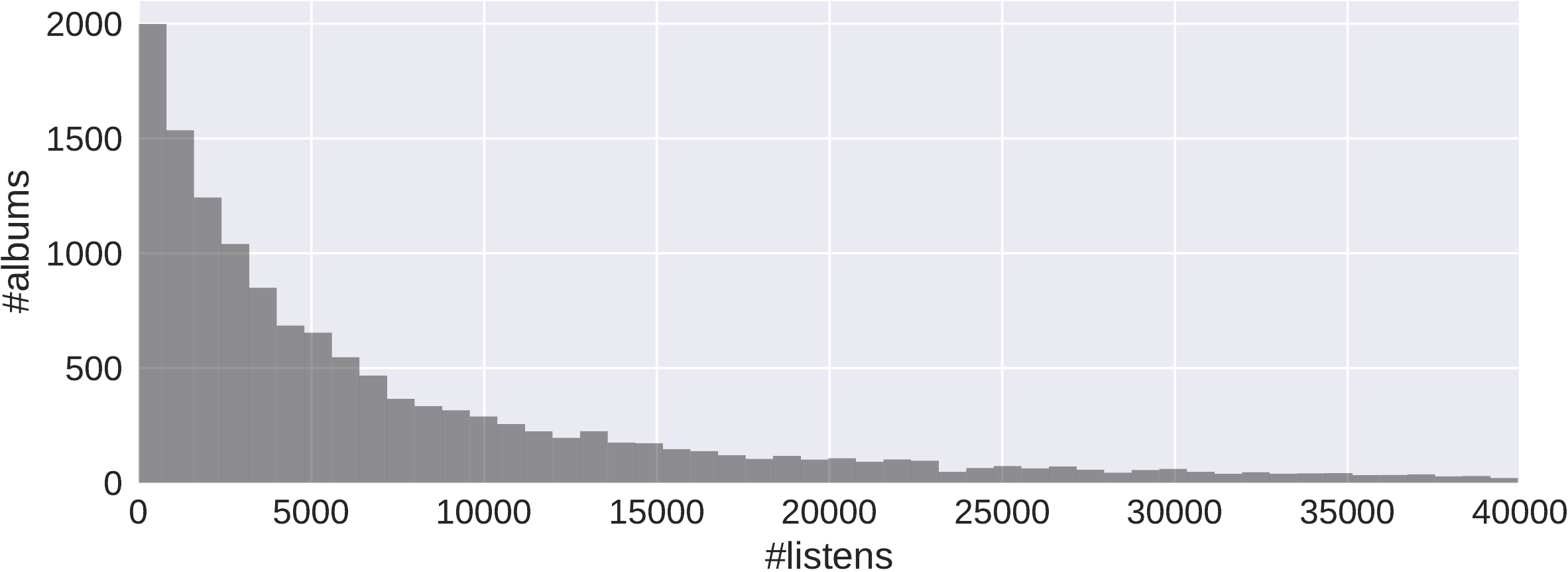}
	\caption{Album listens (min 0, max 3.6 millions).}
	\label{fig:listens_distribution}
\end{figure}

\begin{figure}
	\centering
	\includegraphics[width=\linewidth]{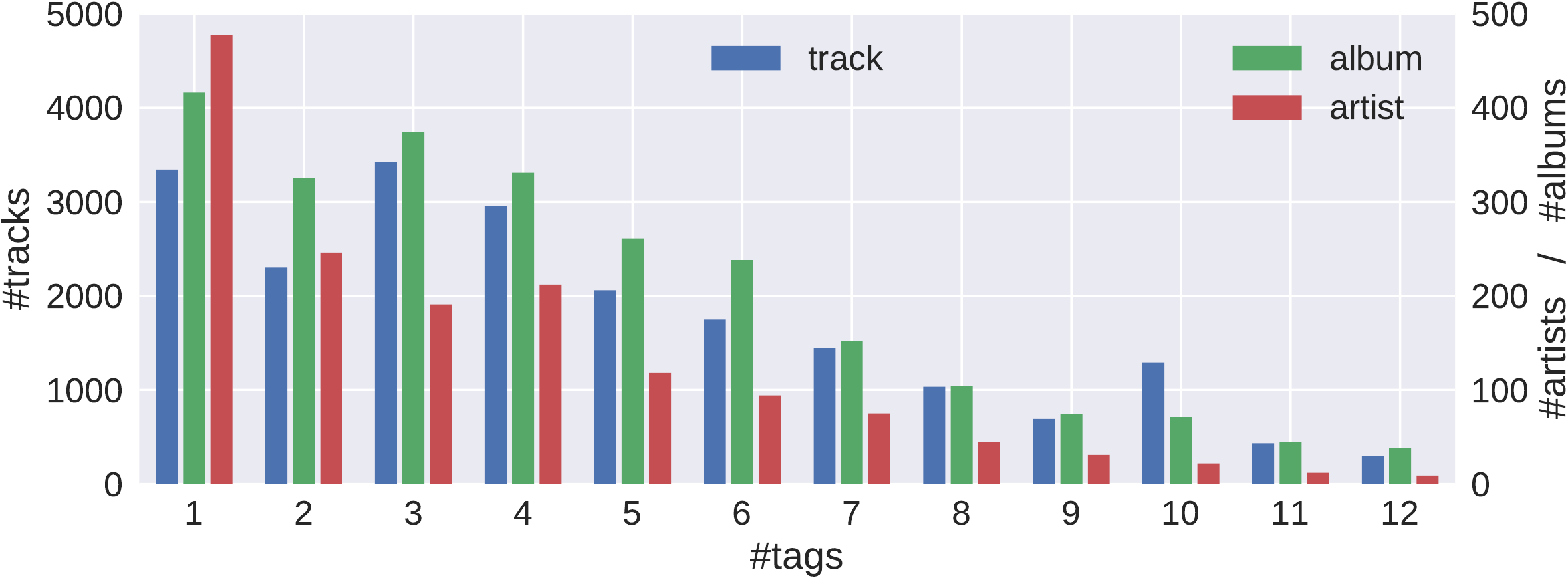}
	\caption{Per-track, album and artist tags (min 0, max 150).}
	\label{fig:tag_distribution}
\end{figure}

\subsection{Content}

The collected metadata\footnote{\texttt{raw\_tracks.csv}, \texttt{raw\_albums.csv}, \texttt{raw\_artists.csv}} was cleaned, uniformly formatted, merged and stored in \texttt{tracks.csv}\footnote{\label{creation}See \texttt{creation.ipynb} for the code which created the dataset.} which \tabref{tab:tracks} shows an excerpt. That file is a relational table where each row represents a track and columns are listed in \tabref{tab:metadata}.
For ease of use, we kept all the metadata in a single table despite the redundancy incurred by the fact that all tracks from a given artist share all artist related columns. The problem is mitigated in practice by compression for storage and by \textit{categorical variables} for memory usage.

All the metadata available through the API has been archived. It includes song title, album, artist, and per-track genres; user data such as per-track/album/artist favorites, play counts, and comments; free-form text such as per-track/album/artist tags, album description and artist biography. 
Coverage varies across fields and is reported in \tabref{tab:metadata}.
Note that all that metadata has been produced by the artists when uploading their music and that while the content is curated, the curators focus on the musical content not the metadata.
Figures~\ref{fig:album_release_year},~\ref{fig:duration_distribution} and \ref{fig:listens_distribution} show the distribution of albums per year, track durations, and play counts per album.
See the \texttt{analysis.ipynb} notebook for a much more detailed analysis of the content.

The audio for each track is stored in a file which name is the track id. All tracks are mp3-encoded, most of them with sampling rate of 44,100 Hz, bit rate 320 kbit/s (263 kbit/s on average), and in stereo.

\begin{table}
	\small
	\centering
	\begin{tabular}{ccclr}
		\toprule
		id & parent & top\_level & title & \#tracks \\
		\midrule
		38 & None & 38 & Experimental & 38,154 \\
		15 & None & 15 & Electronic & 34,413 \\
		12 & None & 12 & Rock & 32,923 \\
		1235 & None & 1235 & Instrumental & 14,938 \\
		\midrule
		25 & 12 & 12 & Punk & 9,261 \\
		89 & 25 & 12 & Post-Punk & 1,858 \\
		1  & 38 & 38 & Avant-Garde & 8,693 \\
		\bottomrule
	\end{tabular}
	\caption{An excerpt of the genre hierarchy, stored in \texttt{genres.csv}. Some of the 16 top-level genres appear in the top part, while some second- and third-level genres appear in the bottom part.}
	\label{tab:genres}
\end{table}

\begin{figure}
	\begin{minipage}[t]{0.44\linewidth}
		\centering
		\includegraphics[width=\linewidth]{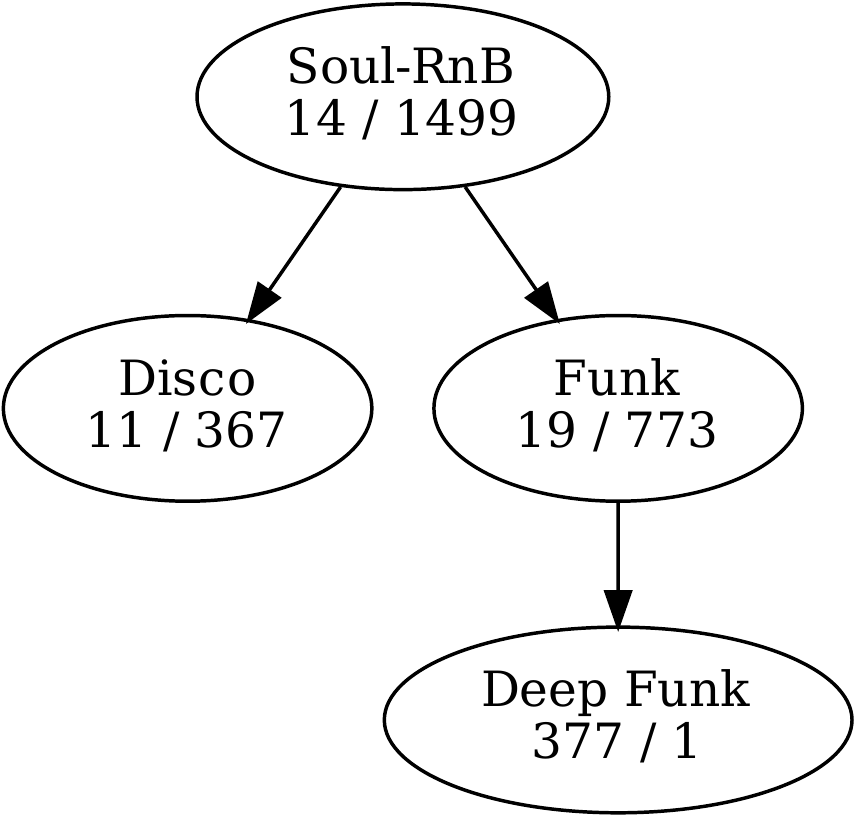}
	\end{minipage} \hfill
	\begin{minipage}[t]{0.55\linewidth}
		\centering
		\includegraphics[width=\linewidth]{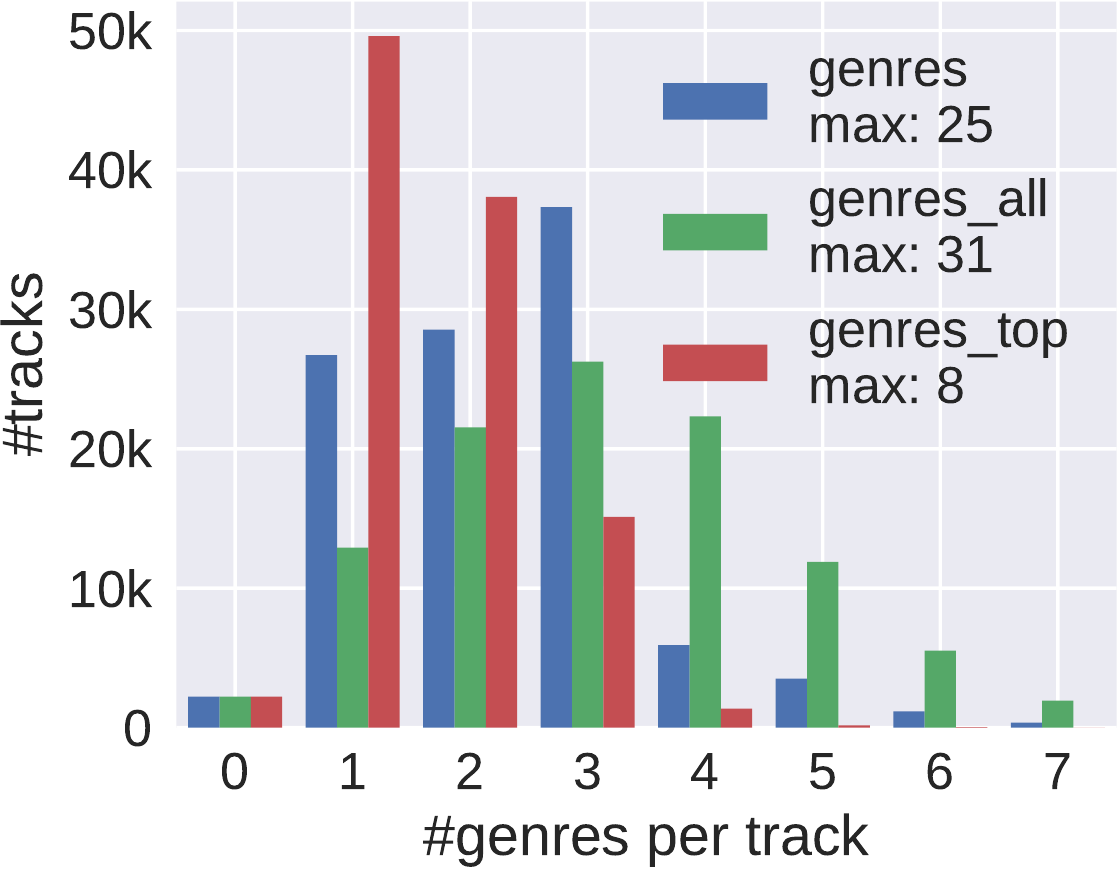}
	\end{minipage}
	\caption{(left) Example of genre hierarchy for the top-level Soul-RnB genre. Left number is the \texttt{genre\_id}, right is the number of tracks per genre. (right) Number of genres per track. A 3 genres limit has been introduced early on by the administrators.}
	\label{fig:genre_hierarchy}
	\label{fig:genres_per_track}
\end{figure}

\subsection{Genres}

The FMA is especially suited for MGR as it features fine genre information, i.e.\ multiple (sub-)genres associated to individual tracks, has a built-in genre hierarchy (\tabref{tab:genres}), and is annotated by the artists themselves. While the artists are the best placed to judge the positioning of their creations, they might be inconsistent and motivated by factors not necessarily objective, such as achieving a higher play count. As labeling noise is unavoidable, those labels should ideally be one of many ground truths, to be complemented by crowd-sourcing and experts (from different music metadata websites).

While there is no agreement on a taxonomy of genres~\cite{mir_review_genre}, we followed the hierarchy used by the archive, which is the one the authors had in mind when annotating their tracks. That hierarchy is composed of \ngenres genres of which 16 are roots, the others being sub-genres. \tabref{tab:genres} shows an excerpt of that information along with the number of tracks per genre and the associated top-level genre, that is the root of the genre tree. \figref{fig:genre_hierarchy} shows an excerpt of the tree.

\begin{figure}[t]
	\centering
	\includegraphics[width=\linewidth]{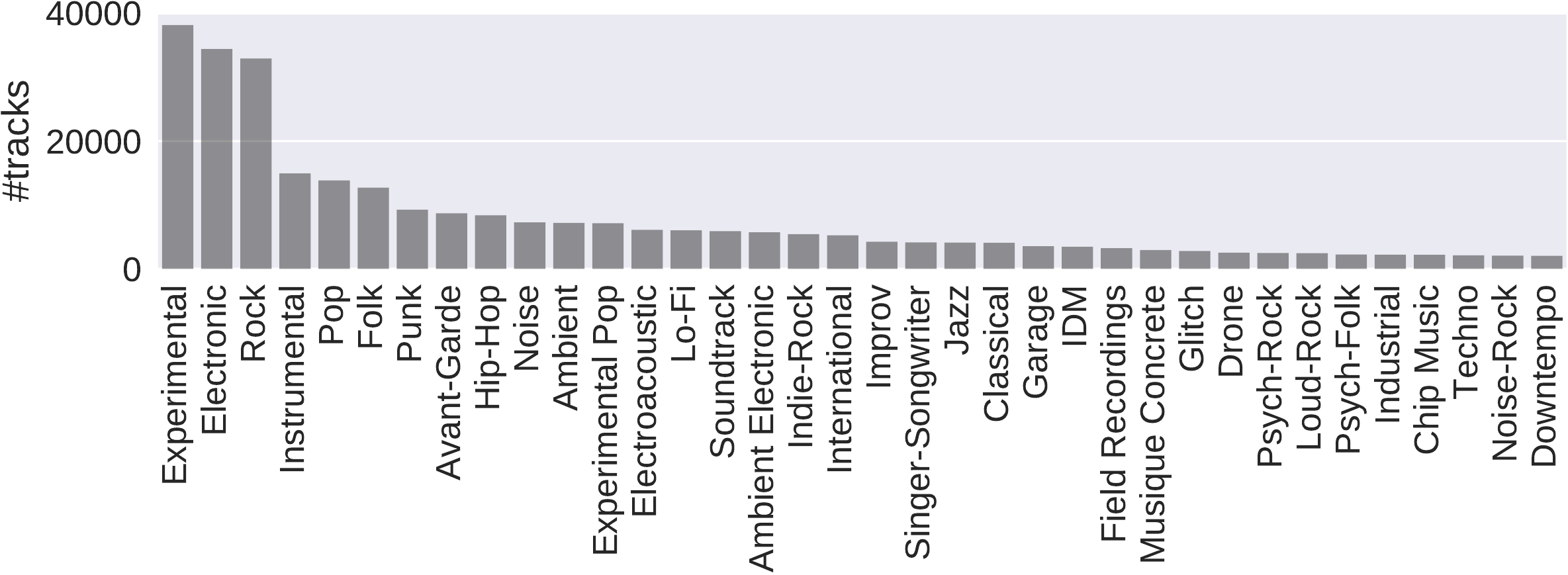}
	\\ \vspace{1em}
	\includegraphics[width=\linewidth]{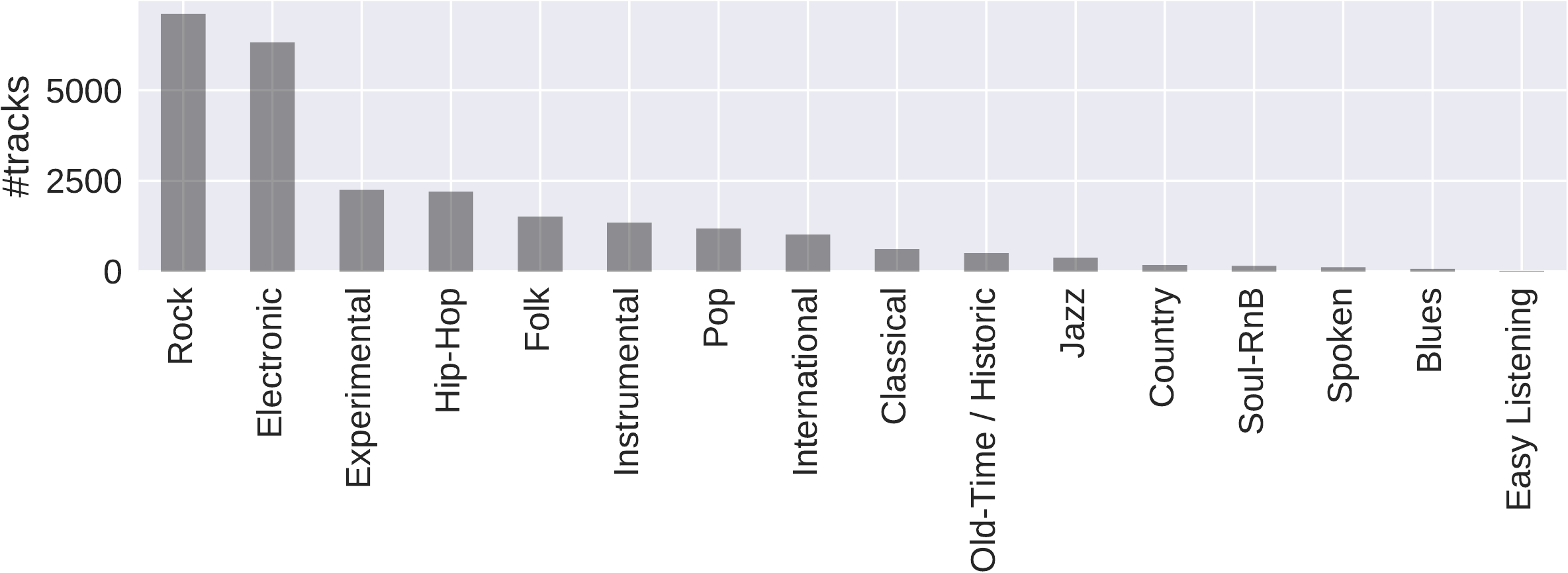}
	\caption{(top) Tracks per (sub-)genre on the full set (min 1, max 38,154). (bottom) Tracks per all 16 root genres on the medium subset (min 21, max 7,103). Note how experimental music is much less represented in the curated medium subset.}
	\label{fig:genre_distribution}
	\label{fig:genre_top_distribution}
\end{figure}

In the per-track table, the \texttt{genres} column contain the genre ids indicated by the artist. Then, given such hierarchical information, we constructed a \texttt{genres\_all} column which contains all the genres encountered when traversing the tree from the indicated genres to the roots.
The root genres are stored in the \texttt{genres\_top} column.
\figref{fig:genres_per_track}~and~\ref{fig:genre_distribution} shows the number of genres per track and tracks per genre.


\subsection{Features} 

To allow researchers to experiment without dealing with feature extraction, we pre-computed the features listed in \tabref{tab:features}. These are all the features the \href{https://github.com/librosa/librosa}{librosa} Python library, version 0.5.0~\cite{librosa}, was able to extract.
Each feature set (except zero-crossing rate) is computed on windows of 2048 samples spaced by hops of 512 samples. Seven statistics were then computed over all windows: the mean, standard deviation, skew, kurtosis, median, minimum and maximum.
Those 518 pre-computed features are distributed in \texttt{features.csv} for all tracks.\footnote{See \texttt{features.py} for the code which computed the features.}


\subsection{Subsets} \label{sec:subsets}

For the dataset to be useful as a development set or for people with lower computational resources, we propose the following sets, each of which is a subset of the larger set:
\begin{enumerate}
	\item \textbf{Full}: the complete dataset, described above. All \ngenres genres, unbalanced with 1 to 38,154 tracks per genre (\figref{fig:genre_distribution}) and up to 31 genres per track (\figref{fig:genres_per_track}).
	\item \textbf{Large}: the full dataset with audio limited to 30 seconds clips extracted from the middle of the tracks (or entire track if shorter than 30 seconds). That trimming reduces the size of the data by a factor 10.
	\item \textbf{Medium}: while root genre recognition should be treated as a multi-label problem in general, we constructed this subset for the simpler problem of single-label prediction. It makes sense as half the tracks have a single root genre (\figref{fig:genres_per_track}). As such, we selected those tracks with only one top genre and sampled the clips according to the completeness of their metadata and their popularity, hoping to select tracks of higher quality. That selection left us with 25,000 30s clips, genre unbalanced with 21 to 7,103 clips per top genre (\figref{fig:genre_top_distribution}), but only one of the 16 top genres per clip.
		\todo{}
	\item \textbf{Small}: to construct a balanced subset, we selected with the same process the top 1,000 clips from the 8 most popular genres of the medium set. The subset is thus composed of 8,000 30s clips from 8 top genres, balanced with 1,000 clips per genre, 1 root genre per clip. This subset is similar to the very popular GTZAN~\cite{gtzan} with the added benefits of the FMA, that is metadata, pre-computed features, and copyright-free audio.
\end{enumerate}
\tabref{tab:subsets} highlights the main differentiating factors between the proposed subsets.

\subsection{Splits}

We propose an 80/10/10\% split into training, validation and test sets to make research on the FMA reproducible. Training and validation shall be merged if cross-validation is used instead.
Below are the followed constraints:
\begin{enumerate}
	\item Stratified sampling to preserve the percentage of tracks per genre (important for minority genres). Each root genre is guaranteed to be represented in all splits, but the ratio is only exact for the small subset (800/100/100). The seven smallest sub-genres (less than 20 tracks in total) are however not guaranteed to appear in all splits of the full and large sets.
	\item An artist filter for artists to be part of one set only, thus avoiding any artist and album effect. It has been shown that the use of songs from the same artist in both training and test sets leads to over-optimistic accuracy and may favor some approaches~\cite{artist_filters_1, artist_filters_2}.
\end{enumerate}
The above constraints are satisfied for all subsets, and a track is assigned to the same split across all of them.\footref{creation} The 2,231 tracks without genre label are assigned to the training set (full and large sets) as they might be useful as additional training samples for semi-supervised algorithms.

\section{Usage} 

With its rich set of metadata, user data, audio and features, the FMA is amenable to many tasks in MIR. We share below some possible uses which serve to illustrate the breadth of data available in the dataset.




\subsection{Music Classification and Annotation}

Music classification is a key problem in MIR with many potential applications. For one, a classification system enables end users to search for the types of music they are interested in. On the other hand, different music types are managed more effectively and efficiently once they are categorized into different groups~\cite{mir_review_classif}.
The classification tasks which can readily be evaluated on FMA include genre recognition, artist identification, year prediction, and automatic tagging.
Automatic tagging~\cite{autotagging} is a classification problem which covers different semantic categories, where tags are labels which can be any musical term that describes the genre, mood, instrumentation, and style of the song.
It helps to convert the music retrieval problem to text retrieval by substituting songs with tags.
In addition to supervised methods which classify music given an arbitrary taxonomy, another approach is to cluster data in an unsupervised way so that a categorization will emerge from the data itself based on objective similarity measures. Then, does genre or another taxonomy naturally come up?

\begin{table}
	\small
	\centering
	\begin{tabular}{lrrccc}
		\toprule
		dataset & clips & genres & length & \multicolumn{2}{c}{size} \\
		\cmidrule{5-6}
		        &       &        &  [s]   & [GiB] & \#days \\
		\midrule
		small  &    8,000 &   8 &  30 &  7.4 & 2.8  \\
		medium &   25,000 &  16 &  30 & 23   & 8.7  \\
		large  & \ntracks & 161 &  30 & 98   & 37 \\
		full   & \ntracks & 161 & \aduration & \size & \tduration  \\
		\bottomrule
	\end{tabular}
	\caption{Proposed subsets of the FMA.}
	\label{tab:subsets}
\end{table}

\begin{table}
	\small
	\centering
	\begin{tabular}{l@{\tspaceb}r@{\tspaceb}c@{\tspaceb}c@{\tspaceb}c@{\tspaceb}c}
		\toprule
		feature set & dim. & LR & kNN & SVM & MLP \\
		\midrule
		1 Chroma~\cite{chroma} & 84 & 44 & 44 & 48 & 49 \\
		2 Tonnetz~\cite{tonnetz} & 42 & 40 & 37 & 42 & 41 \\
		3 MFCC~\cite{mfcc} & 140 & 58 & 55 & 61 & 53 \\
		4 Spec. centroid & 7 & 42 & 45 & 46 & 48 \\
		5 Spec. bandwidth & 7 & 41 & 45 & 44 & 45\\
		6 Spec. contrast~\cite{contrast} & 49 & 51 & 50 & 54 & 53 \\
		7 Spec. rolloff & 7 & 42 & 46 & 48 & 48 \\
		8 RMS energy & 7 & 37 & 39 & 39 & 39 \\ 
		9 Zero-crossing rate & 7 & 42 & 45 & 45 & 46 \\
		\midrule
		3 + 6 & 189 & 60 & 55 & 63 & 54 \\
		3 + 6 + 4 & 273 & 60 & 55 & 63 & 53 \\
		1 to 9 & 518 & 61 & 52 & 63 & 58 \\
		\bottomrule
	\end{tabular}
	\caption{Test set accuracies of various features and classifiers for top genre recognition on the medium subset.}
	\label{tab:mgr}
	\label{tab:features}
\end{table}

\subsection{Genre Recognition} 

Music genres are categories that have arisen through a complex interplay of cultures, artists, and market forces to characterize similarities between compositions and organize music collections. Yet, the boundaries between genres still remain fuzzy, making the problem of music genre recognition (MGR) a nontrivial task~\cite{mir_review_genre}.
While its utility has been debated, mostly because of its ambiguity and cultural definition, it is widely used and understood by end-users who find it useful to discuss musical categories~\cite{mgr_why}.
As such, it is one of the most researched areas of MIR.
We propose the following prediction problems of increasing difficulty:
\begin{enumerate}
	\item Single top genre on the balanced small subset.
	\item Single top genre on the unbalanced medium subset.
	\item Multiple top genres on the large / full set.
	\item Multiple (sub-)genres on the large / full set.
\end{enumerate}

\tabref{tab:mgr} reports accuracies for problem 2 with nine mainstream feature sets and some combinations as well as four standard classifiers using scikit-learn, version 0.18.1~\cite{scikit-learn}. Specifically, we employed linear regression (LR) with an $L^2$ penalty, k-nearest neighbors (kNN) with $k=200$, support vector machines (SVM) with a radial basis function (RBF) kernel and a multilayer perceptron (MLP) with 100 hidden neurons. All classifiers were tested with otherwise default settings.\footnote{See \texttt{baselines.ipynb} for all details.} Reported performance should not be taken as the state-of-the-art but rather as a lower-bound and an indication of the task's difficulty. Moreover, the developed code can serve as a reference and is easily modified to accommodate other features and classifiers.

A major motivation to construct this dataset was to enable the use of the powerful DL set of techniques to music analysis, an hypothesized cause of stagnation on MIREX tasks~\cite{mirex_stagnation}.
With availability of audio, DL architectures such as convolutional neural networks and recurrent neural networks can be applied to the waveform to avoid any feature engineering. While those approaches have fallen behind learning from higher-level representations such as spectrograms~\cite{dieleman_endtoend}, a greater exploration of the design space will hopefully provide alternatives to solving MIR challenges~\cite{mir_dl_feature_learning}.


\subsection{Data Analysis}

While our intention was to release a large volume of audio for machine learning algorithms, analyzing audio is certainly of interest to musicologists and researchers who want to study relations with higher-level representations. 
Moreover, the availability of complete tracks allows proper study of music properties, for example music structure analysis.
Finally, the metadata is surely a valuable addition to existing datasets (e.g. MusicBrainz, AllMusic, Discogs, Last.fm) for metadata analysis.

\section{Discussion} 




While the FMA can be used to evaluate many tasks, metadata is missing for e.g.\ mood classification or instrument recognition. However, a more thorough investigation of the available tags may reveal their feasibilities. Similarly, cover song detection may be doable if multiple versions of many songs are featured.
While the present dump only captures listening and downloading counts in aggregates,\footnote{That information can be useful to e.g.\ analyze and predict hits.} the lists of which songs, albums and artists a user marked as favorites or commented are public, as well as \textit{user mixes}. While not public, listening and downloading activities are logged and might be shared after anonymization.\footnote{Private discussion with the website administrators.} Moreover, users form a public social network via \textit{friend requests}.
Collecting this information would open the possibility of a large-scale evaluation of content-based recommender systems.
Cover images for tracks, albums, and artists are another public asset which may be of interest.
Finally, we can expect the dataset to be cross-referenced with other resources to unlock additional tasks, as has happened for example with the MSD and \href{http://www.allmusic.com}{AllMusic}, \href{https://www.last.fm}{last.fm} and \href{https://beatunes.com}{beaTunes} for genre recognition~\cite{msd_features, msd_genres}, \href{https://musixmatch.com}{musixmatch} for lyrics, \href{https://secondhandsongs.com}{SecondHandSongs} for cover songs, or \href{https://www.thisismyjam.com}{This Is My Jam} for user play counts.

Diversity is another issue. As suggested by \figref{fig:genre_distribution}, this collection is biased toward experimental, electronic, and rock music. Moreover, it does not contain mainstream music and few commercially successful artists.
A common criticism of basing research on CC-licensed music is that the music is of substantially lower ``quality''. Moreover, it is unknown whether datasets made up of mainstream or non-mainstream music have similar properties and if algorithms tailored on one perform similarly on the other. While those points are valid for high-level tasks such as recommendation (which depend on a variety of factors beyond the acoustic content), this is a much more tenuous case for the majority of tasks, in particular perceptual tasks.
Nevertheless, algorithms should ideally be evaluated on multiple datasets, which will help answer such questions.

\section{Conclusion and Perspectives}


Benchmarking is an important aspect in experimental sciences --- results reported by individual research groups need to be comparable. Important aspects of these are datasets that can be easily shared among researchers, together with a set of defined tasks and splits. The FMA enables researchers to test algorithms on a large-scale collection, closer to real-world environments. Even though it is still two orders of magnitude behind commercial services who have access to tens of millions of tracks,\footnote{\href{http://the.echonest.com}{37M Echonest}, \href{https://en.wikipedia.org/wiki/Spotify}{30M Spotify}, \href{http://www.skilledtests.com/wiki/Last.fm_statistics}{45M last.fm}, \href{http://bupz.com/best-websites-to-buy-musics}{45M 7digital}, \href{https://www.apple.com/pr/library/2013/02/06iTunes-Store-Sets-New-Record-with-25-Billion-Songs-Sold.html}{26M iTunes}} it is of the same scale as the largest image dataset which opened the door to dramatic performance improvements for many tasks in computer vision.
By providing audio, we do not limit the benchmarking to pre-computed features and allow scientists to develop and test new feature sets, learn features, or learn mappings directly from the audio.
For now, music classification, and MGR in particular, is the most straightforward use case for FMA. The inclusion of a genre hierarchy makes it specially interesting, as it offers possibilities rarely found in alternative collections.


In addition to the proposed usage and many others people will find, future work on the dataset itself should focus on (i) validating the ground truth by measuring agreement by independent annotators and (ii) obtaining additional metadata and labels.
If the community finds interest in the dataset and validate its use, that can be achieved by scraping the website for information not available through the API,
cross-referencing with other resources, or crowd-sourcing (with e.g.\ \href{https://www.mturk.com}{Mechanical Turk} or \href{https://www.crowdflower.com/}{CrowdFlower}).

In a \href{https://freemusicarchive.org/member/cheyenne_h/blog/FMA_Dataset_for_Researchers}{post about the dataset}, Cheyenne Hohman, the Director at the Archive, wrote that ``by embracing the
~\ldots~
philosophy of Creative Commons, artists are not only making their music available for the public to listen to, but also for educational and research applications''.
Let's hope for a future where sharing is first and researchers feed open platforms with algorithms while they feed us with data.

\section{Acknowledgments}

We want to thank the team supporting the \href{https://freemusicarchive.org}{Free Music Archive} as well as all the contributing artists and curators for the fantastic content they made available.
We want to thank the anonymous ISMIR reviewers for their thorough reviews and many constructive comments which have improved the quality of this work.
We are grateful to SWITCH and EPFL for hosting the dataset within the context of the \href{https://projects.switch.ch/scale-up}{SCALE-UP} project, funded in part by the swissuniversities \href{http://www.swissuniversities.ch/isci}{SUC P-2 program}.
Xavier Bresson is supported by NRF Fellowship NRFF2017-10.

\bibliography{refs}
\end{document}